\documentclass[12 pt,eps]{article}
\usepackage[dvips]{graphicx}

\catcode`\@=11
\def\Bbb{\ifmmode\let\next\Bbb@\else
\def\next{\errmessage{Use \string\Bbb\space only in math mode}}\fi\next}
\def\Bbb@#1{{\Bbb@@{#1}}}
\def\Bbb@@#1{\fam\msbfam#1}

\else
\fi \topmargin -30pt

\ifcase \@ptsize
\or 
\or 
\fi

\def \B{{B\"acklund} }

\def \d{\partial}
\def \db{\bar{\partial}}

\def \dt {\partial_{0}}
\def \dx {\partial_{1}}
\def \s {\sigma_3}

\def \a {{\cal A}}
\def \ab{\bar{{\cal A }}}
\def \b{\beta}
\def \l{\lambda}
\def \p{\phi}
\def \e{\eta}
\def \t{\theta}
\def \vt{\vartheta}
\def \Es {E^{*}}
\def \E {E^{*}}
\def \Ps {P^{*}}
\def \vs {v^{*}}
\def \us {u^{*}}
\def \A{\bar{A}}
\def \J{\bar{J}}

\def \te{\widetilde{E}}
\def \tes{\widetilde{E}^{*}}
\def \tp{\widetilde{P}}
\def \tps{\widetilde{P}^{*}}

\def \tj{\widetilde{J}}
\def \tjb{\widetilde{\J}}
\def \be{\begin{equation}}
\def \ee{\end{equation}}
\def \g{g^{-1}}
\def \x{\chi}
\begin{document}
\title {The complex sine-Gordon model on a half line}
\author{P. Bowcock\footnote{Peter.Bowcock@durham.ac.uk}, G. Tzamtzis\footnote{Georgios.Tzamtzis@durham.ac.uk}}
\maketitle

\bigskip
\centerline{Centre for Particle Theory}
\centerline{Department of Mathematical Sciences}
\centerline{University of Durham} \centerline{Durham DH1 3LE,
England}

\medskip
\vskip 4pc

\begin{abstract}
In this paper, we examine the complex sine-Gordon model in the
presence of a boundary, and derive boundary conditions that preserve
integrability. We present soliton and breather
solutions,  investigate the scattering of particles and
solitons off the boundary and examine the existence of classical solutions corresponding to
boundary bound states.
\end{abstract}

\section{Introduction}
Two dimensional integrable field
theories have become an area of extensive study. Their rich
underlying mathematical structure allows for their exact solution. This in turn provides
valuable information about the wide range of physical phenomena which integrable field theories can be used to model,
and more generally about non-perturbative field theory. A number of models have been studied
 in the presence of a boundary. This 
has led to results for both the
classical and quantum scattering of objects, like particles and solitons, off the
boundary. In particular Toda models
\cite{Bowcock:1995vp}, the sine-Gordon model \cite{Saleur:1995yh},
and the sinh-Gordon model\cite{Corrigan:1998fc} amongst others have
been studied on the half-line.

In this paper we study the classical two dimensional complex sine-Gordon
(CSG) model in the presence of a boundary. The model in the bulk is described by the following Lagrangian

\begin{equation}\label{eq:CSGlagr2}
{\cal L}_{CSG}  =  \frac{1}{2}\frac { {\d u} {\db \us} +{\db u}
{\d \us}}{1 - \xi u u^{*}}- 4 \beta u u^{*} \ ,
\end{equation}
where $ \d = \d / \d z $,  $\db = \d/\d \bar{z}$, $z= (t-x)/2$,
$\bar{z} = (t+x)/2$ . The field $u$ is complex, $\b$ is a real
coupling constant and the parameter $\xi$ can be changed by
rescaling $u$ and will henceforth be set to one. The Lagrangian
is invariant under global phase rotations of the field $u$ and this
leads to a conserved $U(1)$ charge.

As such, the complex sine-Gordon model comprises an integrable
generalisation of the sine-Gordon theory, with an internal $U(1)$
degree of freedom. It was first introduced independently by Lund
and Regge as a model of relativistic vortices in a superfluid
\cite{Lund:1976ze,Lund:1977dt} and by Pohlmeyer in a dimensional
reduction of a O(4) non-linear $\sigma$-model
\cite{Pohlmeyer:1975nb}.
It belongs to a class of generalisations of the
sine-Gordon theory, which in the literature is referred to as
homogeneous sine-Gordon theories. The latter appear as $G/U(1)$
gauged Wess-Zumino-Witten models perturbed by a potential. These
generalisations  describe integrable perturbations of $c>1$
conformal field theories and have been studied in the bulk in
\cite{Fernandez-Pousa:1997hi,Fernandez-Pousa:1998iu}. The CSG
model appears as the simplest case where $G=SU(2)$ and describes
integrable perturbations of {\bf Z}$_k$ parafermions by the first
thermal operator \cite{Bakas:1994xh,Park:1994bx}. The quantum case
was also studied by Dorey and Hollowood in \cite{Dorey:1995mg},
and Maillet and de Vega in \cite{deVega:1983sh}.

The CSG model has found many applications in different fields of
physics, from general relativity \cite{deVega:1994pm}, to the
description of propagating optical pulses in a non-linear medium
\cite{Park:1995ya}. In the latter, the CSG theory was used as a
generalisation of the pioneering work of McCall and Hahn
\cite{McCall:1967}, where the simple sine-Gordon (SG) was used as
a field theory description. The CSG theory provides for a more
realistic description of optical pulses incorporating effects like
frequency detuning and modulation, while at the same time the
theory can be extended beyond the description of two-level atom systems to
multi-level atom systems.

In the next section of this paper some aspects of the theory in
the bulk will be reviewed. Classical solutions, including
particles, solitons and breathers will be presented for the
model in the bulk clarifying previous
treatments. Unlike the sine-Gordon model where the soliton
solutions are topological in nature,  the CSG model solitons which
carry a N\"oether $U(1)$ charge, have no obvious topological
charge associated with them. This follows from the trivial vacuum
structure of the model for $\b>0$, although as we shall explain
chargeless solitons do carry the topological charge of the
sine-Gordon model in a subtle way. We also discuss the existence
of breathers and find a class of charged breather solutions.

In section 3, we consider the effect of introducing a boundary and 
suitable boundary
conditions are found which preserve the integrability of the
model.In order to do this we explicitly construct low-spin conserved
charges by abelianising the Lax pair description of the model, and ensure
that these are conserved on the half-line by imposing boundary conditions.
We find a natural generalisation of the boundary conditions that Ghoshal 
and Zamolodchikov \cite{Ghoshal:1994tm}
introduced for the sine-Gordon theory.

The final section in the main body of the paper deals with
the scattering of particles and solitons. We find the corresponding
solutions and use them to calculate the classical time-delay.
We also find solutions corresponding to boundary bound states.
The paper
concludes with some general remarks about the results presented,
as well as a few open questions that are related with some
interesting properties of the model.

\newpage
\section{Classical aspects of the CSG theory}
In this section we shall review some important features of the CSG
theory and establish our notation. We present the equation of
motion, the relation of the model with the sine-Gordon theory and
review the mathematical background in which the CSG model can be
regarded as a perturbed gauged Wess-Zumino-Witten model. Moreover vacuum,
soliton and multi-soliton solutions are written down in a compact
form. Finally, we clarify the existing confusion concerning the
existence of breathers within this model and present explicit
breather solutions.

\subsection{Definition of the model}
The Lagrangian of the model was presented in (\ref{eq:CSGlagr2}).
We assume  without loss of generality $\xi=1$ since the parameter
can be dropped by rescaling the field variables.  The equation of
motion that follows  is
\begin{equation}\label{eq:CSGeq1u}
\d \db u + \frac{\us \d u \db u}{1- u \us} + 4 \beta u (1- u \us)
= 0 \ \ .
\end{equation}
The relation between this model and sine-Gordon becomes obvious if
we substitute
 \be \label{eq:u2phi}
 u =\sin \p \ e^{2 i \e} \ ,
 \ee
with $\p$ and $\e$ real fields. We use this notation here which
gives a more standard connection with the sine-Gordon theory but
differs from previous treatments by  $\p \rightarrow
\frac{\pi}{2}- \p$. This takes (\ref{eq:CSGlagr2}) to
 \be \label{eq:Lagphi}
{\cal{L}} = \d \p \db \p + 4 \tan^2 \p \d \e \db \e -4 \beta
\sin^2 \p \ .
 \ee
This is essentially the form of the Lagrangian derived by Lund and
Regge, and Pohlmeyer. By taking the field $\e$ to be constant the
sine-Gordon Lagrangian emerges. As was demonstrated by Bakas
\cite{Bakas:1994xh} the theory can be reformulated in terms of a
gauged WZW action. The corresponding action principle can be
written as

 \be \label{eq:wzw1}
 S=S_{gWZW}+S_{pot} \ \ .
 \ee
The action term $S_{gWZW}$ is the well known gauged WZW action
 \begin{eqnarray}
 S_{WZW} &=& -\frac{1}{4 \pi} \int_{\Sigma} dz d\bar{z}
  \ \mbox{Tr} (\g \d g  \g \db
 g) - \frac{1}{12 \pi} \int_B \ \mbox{Tr} (
 \tilde{g}^{-1}d\tilde{g} \wedge \tilde{g}^{-1}d\tilde{g} \wedge
 \tilde{g}^{-1}d\tilde{g}) \nonumber \\ & & + \frac{1}{2 \pi}
 \int \ \mbox{Tr}(- W \db g \g +\bar{W}
\g \d g + W g \bar{W} \g -W \bar{W}) \ \ .
  \end{eqnarray}
This action is defined in a three-dimensional manifold $B$ whose
boundary is our compactified normal two-dimensional space
$\Sigma$. The field $g$ is an $SU(2)$ group element and
$\tilde{g}$ is the extension of $g$ to the three dimensional
manifold. The last term introduces gauge fields $W$ and $\bar{W}$
which act as Lagrange multipliers. The $S_{pot}$ term is

\be
 S_{pot} = \frac{\beta}{2\pi} \int \  \mbox{Tr}  (g \sigma_3 g^{-1}
 \sigma_3) \ \ .
 \ee
This  term breaks conformal invariance and thus gives rise to
massive states. Varying the action yields the CSG equations of
motion which can be expressed in a zero curvature form
\cite{Park:1996rj}

\be \label{eq:Lax} [ \d + (\g \d g + \g W g +i \beta \lambda
\sigma_3) \ , \ \db + (\bar{W}-\frac{i}{\lambda} \g \sigma_3 g) \
] = 0 \ .
 \ee
From the variation of the gauge fields $W$ and $\bar{W}$, two
constraint equations arise
 \begin{eqnarray} \label{eq:const2}
 \db g \g - g \bar{W} \g +\bar{W}  = 0  \ , \nonumber \\
 \\
 \g \d g + \g W g - W = 0 \ , \nonumber
 \end{eqnarray}
which are critical in order to make the identification with the
CSG theory. The connexion between the $SU(2)$ matrix $g$ and the
complex field $u$ of (\ref{eq:CSGeq1u}) is given by
 \be\label{eq:g}
 g = \left( \begin{array}{rr}
 u & -i \vs \\
 -i v & \us \
 \end{array} \right) \ ,
 \ee
where $ v=-\sqrt{1-u \us} e^{-i \theta}$ . The field variable $\t$
should not be considered as an independent field but rather as an
auxiliary field that is properly defined up to a constant through
the constraint equations (\ref{eq:const2}). In the gauge where $W
= \bar {W} = 0$ , the constraint equations take the form

\be
 \d \t  = -i \frac{ \us \d u - u \d \us }{2 (1- u \us)} \ \ , \ \
 \db \t =  -i \frac{ u \db \us - \us \db u }{2 (1- u \us)} \ \ ,
 \ee
whilst the equation of motion now becomes

\be \label{eq:lax2}
 [ \d - A , \db - \A ] = 0  \ ,
 \ee
where \be \label{eq:A} A= - (\g \d g + i \beta \lambda \sigma_3) \
\ , \ \ \A = \frac{i}{\lambda} \g \sigma_3 g \ \ .
 \ee

This compact zero-curvature form of the equations of motion
demonstrates the integrability of the model and will prove useful
when we come to consider \B transformations and conserved
quantities in later sections.

\subsection{Vacuum solutions in the bulk.}
From  (\ref{eq:CSGlagr2}) it is easy to see that the energy of the
CSG model in the bulk is
 \be \label{eq:bulkenergy}
 {\cal H}_{bulk} = \int dx \left(\frac{|\dt u|^2 +|\dx u|^2}{1 - u \us}
 + 4 \b u \us \right) \ \ .
 \ee
The most suitable candidate for a vacuum, would be a constant
value for the field $u$ that would force the kinetic term
involving derivatives to vanish and at the same time minimize the
potential term. It is clear to see from (\ref{eq:u2phi}) and
(\ref{eq:Lagphi}) that for $\b>0$ the obvious choice is $u=0$,
while for $\b<0$ the choice should be $|u| = 1$ if we insist that
$|u|\leq 1$ . The sign of the parameter $\b$ divides the theory
into two sectors. In the matrix potential formalism, both sectors
are treated simultaneously as the diagonal and off-diagonal parts
of the field variable $g$. In this context the fields $u$ and $v$
are both solutions to the CSG equation each derived for a specific
choice of $\b$, and each corresponding to a different vacuum .
This is because the two sectors are connected by a duality
transform which interchanges the sign of the coupling constant
$\b$ \cite{Park:1995gc} and simultaneously interchanges the role
of $u$ and $v$ . That is to say the theory is invariant under the
change
 \be \label{eq:dual1}
 g \rightarrow g' = i \sigma_1 g = \left( \begin{array}{cc}
   v & i \us \\
   i u & \vs \
 \end{array} \right) \ \ , \ \ \b \rightarrow - \b ,
 \ee
the latter representing a transform akin to the Krammers-Wannier
duality of the $Z_n$ parafermion theory \cite{Knizhnik:1984nr}.
Taking into account the invariance of the theory under this
duality transform, we shall concentrate in this paper on the $\b
> 0 $ sector which corresponds to the diagonal part of the matrix
formalism. A suitable vacuum solution would be
 \be \label{eq:vacuum}
 g_{vac} = \left( \begin{array}{cc}
   0 & i e^{- i \Omega} \\
   i e^{i \Omega} & 0 \
 \end{array}\right) \ .
 \ee
This selection is consistent with the choices appearing in the
beginning of this section with the diagonal $\b>0$ sector,
corresponding to the $u=0$ vacuum , while the off-diagonal $\b<0$
to $|v|=1 $. It is noted that the apparent singular behaviour of
the Lagrangian at $|u|=1$ does not appear as a problem embedded in
the theory but is a direct consequence of the fact that the gauge
fields $W,\bar{W}$ are ill defined at the specific point.

\subsection{Spectrum of the model}
The CSG model, like the sine-Gordon theory, possesses both
particle and soliton solutions. When small perturbations around
the vacuum are considered
 \be
 u = 0 + \epsilon(x,t) \ ,
 \ee
the theory becomes linear when higher order terms in $\epsilon$
are ignored
\begin{eqnarray}
(\partial^2_0 - \partial^2_1) \epsilon(x,t) + m^2 \epsilon(x,t) =
0 ,
\end{eqnarray}
where $m^2= 4 \b$. The solution to the above equation is the
familiar plane waves solution
 \be \label{eq:planewave}
 \epsilon(x,t) = e^{-i \omega t} \left ( A
  e^{ikx} + B e^{-ikx} \right) ,
 \ee
where $k$ and $\omega$ are related through
\begin{eqnarray}
\omega^2 = k^2 + m^2 .
\end{eqnarray}

Different techniques have been used for the construction of
soliton solutions like the inverse scattering method
\cite{deVega:1983sh} and the Hirota method \cite{Getmanov:1977}.
However both methods yield results that are both cumbersome and
difficult to manipulate. The \B transformation for the CSG model
provides a more elegant way to obtain soliton solutions and can be
written in terms of two matrix variables $g$ and $f$
\cite{Park:1995gc}

\begin{eqnarray}
 \g \d g - f^{-1} \d f - \frac{ \delta \b }{\sqrt{|\b|}}[ \g \s f ,
 \s] &=&0 \ ,\\
 \db g \g \s - \s \db f f^{-1} + \frac{\sqrt{|\b|}}{\delta}(g
 f^{-1} \s - \s g f^{-1}) &=& 0 \ .
\end{eqnarray}
It is easy to show that both $f$ and $g$ satisfy the CSG equation
as well as the constraint equation in the specific gauge choice.
Taking $f$ to be an already known solution, one can generate a new
solution through the equations presented above. One-soliton
solutions can be derived by applying the \B transformation on the
vacuum solutions $g_{vac}$ of (\ref{eq:vacuum}). Each sector of
the theory provides us with two sets of two first order
differential equations that can be integrated, in order to provide
the one-soliton solutions. The diagonal elements of $g$ which
correspond to the $\b>0$ sector give
\begin{eqnarray}\label{eq:back1}
 \dt u - \sqrt{\b} \left( \delta \ e^{i(\t + \Omega)}
 - \frac{1}{\delta} \  e^{-i(\t + \Omega)} \right) \ u \sqrt{1 - u \us}
 &=& 0 \ \ ,\nonumber \\  \\
 \dx u + \sqrt{\b} \left( \delta \ e^{i(\t + \Omega)}
 +\frac{1}{\delta} \  e^{-i(\t + \Omega)} \right) \ u \sqrt{1 - u \us} &=& 0  \ \ . \nonumber
\end{eqnarray}
The one-soliton solution that emerges is
 \be \label{eq:Usoliton}
 u = \frac{\cos(a) \exp \left(2 i \sqrt{\b} \sin(a) \ \frac { t - V x
 }{\sqrt{1 - V^2}}\right)}{\cosh \left({2 \sqrt{\b} \cos(a) \frac{ x - V
 t}{\sqrt{ 1- V^2}}}\right)} \ ,
 \ee
where $V$ and $a$ are real parameters associated with the velocity
and charge of the soliton respectively. This solution was
originally derived by Getmanov \cite{Getmanov:1977} for the $\b
> 0$ case. In addition an expression for the phase $\t$ which
appears in its dual field $v$, is also obtained
 \be \label{eq:theta}
 \theta = -
 \Omega - \arctan \left( \tan(a) \coth \left( 2 \sqrt{\beta}
 \cos(a)
 \frac{ x - V t}{\sqrt{1-V^2}} \right) \right) \ .
 \ee
Respectively for the off-diagonal elements that correspond to the
$\b<0$ sector, the set of equations is
\begin{eqnarray}\label{eq:back2}
 \dt v - \sqrt{|\b|} \ e^{i \Omega} \left( \delta - \frac{1}{\delta}
 \right) ({1 -
 v
 \vs})
 &=& 0 \nonumber \\  \\
 \dx v + \sqrt{|\b|} \ e^{i \Omega} \left( \delta + \frac{1}{\delta}
  \right) ({1 -
 v \vs})
 &=& 0 \ , \nonumber
\end{eqnarray}
that finally produce a different solution
 \be \label{eq:Vsoliton}
 v = - e^{i \Omega} \left( \cos(a) \tanh \left( 2 \sqrt{|\b|}
 \cos(a) \frac{ x - V t}{\sqrt{1-V^2}} \right) + i \sin(a) \right)
 ,
 \ee
with $\Omega$ a real parameter associated with the vacuum of the
theory. This is the solution that was derived by Lund and Regge
\cite{Lund:1976ze} when considering the $\b < 0$ case.

A two-soliton solution can be obtained through a non-linear
superposition technique. Starting from the vacuum of the theory
and by the application of the \B transformation twice, a set of
parameters \{$\delta_1,\delta_2$\} is used respectively in each
step. The same procedure is followed again where the two
parameters are used in the opposite order. By demanding that the
two results are equal, one ends up with an equation for the
two-soliton solution in matrix form
 \be
 g_{2s} = \sigma_3 \left( \delta_1 g_2 - \delta_2 g_1 \right) g_{vac}
 \sigma_3 \left( \delta_1 g_1^{-1} - \delta_2 g_2^{-1}
 \right)^{-1} \ .
 \ee
The matrix field variables $g_k$ are of the general form of
(\ref{eq:g}), with elements
 \be \label{eq:1soliton}
 u_k = \frac{ \cos{(a_k)} N_k \exp(2 i \sqrt{\b} \sin{(a_k)} \Theta_k)}
 {\cosh{\left(2 \sqrt{\b} \cos{(a_k)}
 \Sigma_k
 \right)}
 } \ ,
 \ee
 \be
 v_k = -e^{i \Omega} \left( \cos{(a_k)}  \tanh{\left(2 \sqrt{\b}
 \cos{(a_k)} \Sigma_k\right)} + i  \sin{(a_k)} \right) \ .
 \ee
The identification one must make is:
\begin{eqnarray}
\Sigma_k = \frac{1}{2}\left(\delta_k +\frac{1}{\delta_k}\right) x
+\frac{1}{2}\left (\delta_k - \frac{1}{\delta_k}\right) t \ \ ,
\\
\Theta_k = \frac{1}{2}\left(\delta_k +\frac{1}{\delta_k}\right) t
+\frac{1}{2}\left (\delta_k - \frac{1}{\delta_k}\right) x \ \ ,
\end{eqnarray}
where $N_k$ is a total phase. As expected $g_{2s}$ has the same
general form of equation (\ref{eq:g})
\begin{eqnarray}
g_{2s}=\left( \begin{array}{rr} u_{2s} & -i v_{2s}^* \\ -i v_{2s}
& u_{2s}^*
\end{array} \right ) \ \ .
\end{eqnarray}
The two-soliton solution and its complex conjugate are given by
the diagonal elements of $g_{2s}$
 \be \label{eq:u2s}
 u_{2s}={\frac {\left (-\delta_{{1}}{\it
 \vs_2}+\delta_{{2}}{\it \vs_{1}} \right
 ){e^{i\Omega}}\left (\delta_{{1}}u_{{1}}-\delta_{{2}}u_{{2}}
 \right )+\left (-\delta_{{1}}u_{{2}}+\delta_{{2}}u_{{1}} \right
 ){e^{-i\Omega}}\left (-\delta_{{1}} v_{{1}}+\delta_{{2}}v_{{2}}
 \right )}{{\delta_{{1}}}^{2}+\left (-{\it
 \us_{1}}u_{{2}}-{\it \us_{2}}u_{{1}} -{\it
 \vs_{1}}v_{{2}}-{\it \vs_{2}}v_{{1}}\right
 )\delta_{{2}}\delta _{{1}}+{\delta_{{2}}}^{2} }} \ ,
 \ee
 \be \label{eq:ub2s}
 \us_{2s}={\frac {\left (-\delta_{{1}}{\it
 {v}_2}+\delta_{{2}}{\it {v}}_{{1}} \right
 ){e^{-i\Omega}}\left (\delta_{{1}}\us_{{1}}-\delta_{{2}}\us_{{2}}
 \right )+\left (-\delta_{{1}}\us_{{2}}+\delta_{{2}}\us_{{1}} \right
 ){e^{i\Omega}}\left (-\delta_{{1}} \vs_{{1}}+\delta_{{2}}\vs_{{2}}
 \right )}{{\delta_{{1}}}^{2}+\left (-{\it
 \us_{1}}u_{{2}}-{\it \us_{2}}u_{{1}} -{\it
 \vs_{1}}v_{{2}}-{\it \vs_{2}}v_{{1}}\right
 )\delta_{{2}}\delta _{{1}}+{\delta_{{2}}}^{2} }} \ ,
 \ee
while the off diagonal elements represent the dual field and its
conjugate
 \be
  v_{2s}={\frac {\left (-\delta_{{1}}{\it
 \us_{2}}+\delta_{{2}}{\it \us_{1}}\right
 ){e^{i\Omega}}\left (\delta_{{1}}u_{{1}}-\delta_{{2}}u
 _{{2}}\right )-\left (-\delta_{{1}}v_{{2}}+\delta_{{2}}v
 _{{1}}\right ){e^{-i\Omega}}\left
 (-\delta_{{1}}v_{{1}}+\delta_{{2}}v_ {{2}}\right
 )}{{\delta_{{1}}}^{2}+\left (-{\it \us_{1}}u_{{2}}-{\it
 \us_{2}}u_{{1}} -{\it \vs_{1}}v_{{2}}-{\it
 \vs_{2}}v_{{1}}\right )\delta_{{2}}\delta
 _{{1}}+{\delta_{{2}}}^{2} }} \ ,
 \ee
 \be
  \vs_{2s}={\frac {\left (-\delta_{{1}}{\it
 {u}}_{{2}}+\delta_{{2}}{\it {u}}_{{1}}\right
 ){e^{-i\Omega}}\left
 (\delta_{{1}}\us_{{1}}-\delta_{{2}}\us
 _{{2}}\right )-\left
 (-\delta_{{1}}\vs_{{2}}+\delta_{{2}}\vs
 _{{1}}\right ){e^{i\Omega}}\left
 (-\delta_{{1}}\vs_{{1}}+\delta_{{2}}\vs_ {{2}}\right
 )}{{\delta_{{1}}}^{2}+\left (-{\it \us_{1}}u_{{2}}-{\it
 \us_{2}}u_{{1}} -{\it \vs_{1}}v_{{2}}-{\it
 \vs_{2}}v_{{1}}\right )\delta_{{2}}\delta
 _{{1}}+{\delta_{{2}}}^{2} }} \ .
 \ee
The expressions above represent two-soliton solutions to the
equation of motion and are related through the duality
transformation of (\ref{eq:dual1}).

Multi-soliton solutions can also be obtained by following the same
technique. Instead of the vacuum solution, one can start from any
given n-soliton solution $S_n$ and add solitons through the method
described above, ending up with a $S_{n+2}$ solution.
Nevertheless, multi-soliton solutions for this model are quite
large and their calculation is beyond the scope of this paper.

\subsection{Soliton - Antisoliton duality} \label{sec:sad}

In this section we argue that soliton-soliton and
soliton-antisoliton solutions presented in the literature by
previous treatments, do not represent distinct classes of
solutions. Charged solitons are non-topological solutions,
therefore a distinction between a soliton and an antisoliton is
impossible. On the other hand chargeless solutions may be realised
as topological solitons and identified with the sine-Gordon
solitons.

\begin{figure}[ht]
\begin{center}
\fbox{
\includegraphics[width=.75\textwidth,height=.5\textwidth]{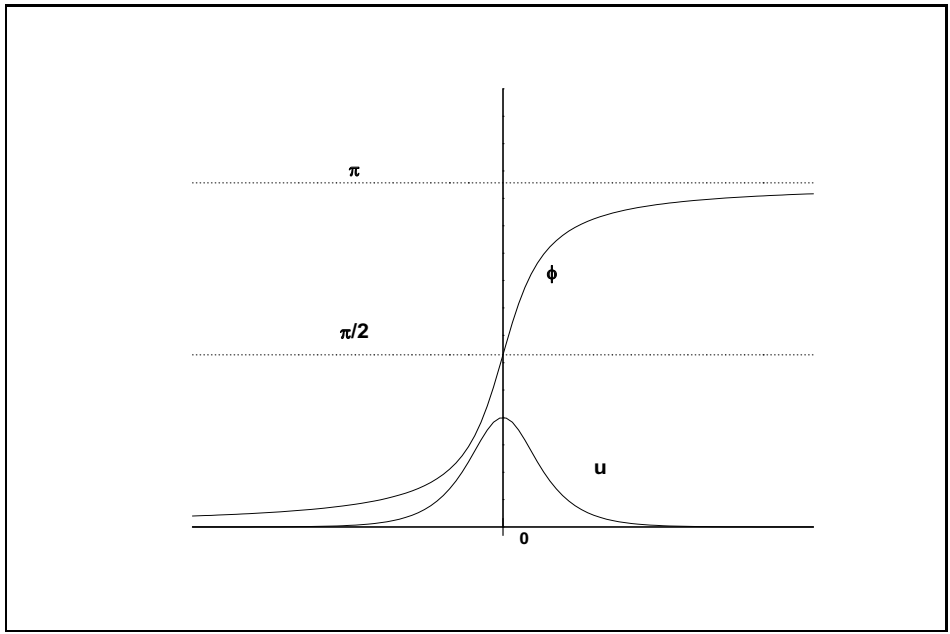}}
\end{center}
\caption{The $\p$ and $u$ solitons}\label{fig:vacuum}
\end{figure}

The sine-Gordon theory appears as the limit of the CSG model when
the charge parameter $a$ is set to zero. We can substitute in the
equation of motion of (\ref{eq:CSGeq1u})
 \be \label{eq:SG}
 u=  \sin\phi e^{2i\e}\ \ ,
  \ee
where $\e$ is now a constant to recover the sine-Gordon model in
the usual form
 \be
 \dt^2 \p - \dx^2 \p + 2 \b \sin 2 \p = 0 \ \ .
 \ee
The sine-Gordon theory has topological solitons (both kinks and
antikinks) interpolating between its degenerate vacuaa. In
contrast the CSG theory has a single vacuum for $\b>0$, and
therefore its solitons are not topological in nature, but are
stable because of integrability alone. The topological nature is
hidden within the mapping of (\ref{eq:SG}) and one has to be
careful when trying to recover the sine-Gordon soliton as a limit
of the CSG theory. Nonetheless a subtle remnant of the topology
survives the mapping to the complex sine-Gordon theory. To see
this consider how a SG soliton is mapped to CSG soliton. This is
shown in Fig.\ref{fig:vacuum}.

Consider now how the potential term behaves as $x$ increases for
the single soliton solution. As a function of $u$ we can express
the potential as
 \be
 \sin 2 \p = 2 \sin \p \cos \p = \pm 2 u \sqrt{1- u \us} \ \ ,
 \ee
where $\e$ has been ignored as a total phase. Note that we should
take the branch cut with opposite signs on each side of the point
$\p=\frac{\pi}{2}$, $u=1$. We shall see that the changing sign of
the branch cut for a chargeless soliton will be important when we
come to consider the theory with a boundary.

In some sense the topology of the chargeless $u$-soliton is
embedded in the the branch cut that appears at the singular point
$u=1$. The choice of branch corresponds to a different  vacuum for
$\p$ and therefore to a different topological charge.

However, when the charge parameter $a$ is not zero, then the
$u$-soliton does not reach the sick point $u=1$ and remains
non-topological. In this case no real distinction can be made
between a soliton and an antisoliton. In the sine-Gordon case, the
antikink solution is derived from the kink by changing the sign of
the parameter $\delta$ of the \B transformation. This effectively
corresponds to a parity and time reversal transformation which
finally produces an antikink solution. Examining
(\ref{eq:1soliton}) we see that in the CSG case this change
actually leads to the complex conjugate solution, by changing the
sign of the complex phase. The change of sign in both $t$ and $x$,
can be cancelled by taking the charge parameter $a\rightarrow -a$.
It is thus clear that instead of changing the sign of $\delta$,
one could effectively change the sign of $a$ to derive an
antisoliton. Since the soliton solution is a smooth function in
$a$, the antisoliton is not a distinct object but can be
identified with the soliton itself.

This also has an effect on the two-soliton solution. If we follow
the same steps as in the sine-Gordon two-soliton solution then the
solution $u_{2s}$ of (\ref{eq:u2s}) corresponds to both a
soliton-soliton and a soliton-antisoliton solution depending on
the choice of sign for the \B parameter $\delta_2$. The parameter
$\delta_2$ can be chosen in such a way as to describe one of the
following
\begin{eqnarray} \label{eq:delta}
\delta_1 &= -(\delta_2)^{-1}&=\sqrt{\frac{1-V}{1+V}} \ \ \ \ \
\mbox{soliton-soliton scattering }\\  \delta_1 &=
(\delta_2)^{-1}&=\sqrt{\frac{1-V}{1+V}} \ \ \ \ \ \mbox{
soliton-antisoliton scattering}
\end{eqnarray}
Here we have taken the two solitons to have equal and opposite
velocity (In general this is not the centre of mass since
differently charged solitons have different masses, but it will be
convenient for our discussion when we introduce a boundary later
on). Also for reasons of simplicity we will refer to the
soliton-soliton solution as $u_{ss}$ ( $ \delta_2 = -1 / \delta_1
$ ) and to the soliton-antisoliton as $u_{sa}$ ($\delta_2 =1/
\delta_1$).

However since no topological distinction exists between the
soliton and antisoliton sector, it is possible to find a
transformation of the parameters of the solution which effectively
acts as a change of sign for the parameter $\delta_2$. In fact, a
set of transformations exists that maps $u_{ss}$ to $u_{sa}$ but
we restrict ourselves to the simplest cases.

Before introducing the transformation, we need to introduce
arbitrary shifts in $x$, which are crucial not only for this
mapping but also later when we consider breathers and soliton
reflections. The shifts appear in exponentials, so it is more
helpful to consider the shifts in the following forms
\begin{eqnarray} \label{eq:shifts}
K_i &=& \exp \left(2 \sqrt{\beta} \cos{a_i}
\frac{x_i}{\sqrt{1-V^2}} \right)  \ \nonumber
\\ J_i &=& \exp \left(2 i \sqrt{\beta} \sin{a_i}
\frac{V y_i}{\sqrt{1-V^2}}+i R_i\right);\ \ \ i
 = 1 ,2 \ .
\end{eqnarray}
The parameters $K_i,J_i$ are directly related with both $V$ and
$a$ and correspond to the arbitrary initial positions in $x$, in
the real ($\Sigma_i$) and imaginary phases ($\Theta_i$)
respectively, that appear in the one-soliton solution.
Specifically the parameter $K$ represents a translation in $x$,
while the $J$ parameter represents a phase shift in the internal
$U(1)$ space. For reasons of simplicity, we include in the
definition of $J$ the total phase $N_k=\exp(i R_k)$ which appears
in (\ref{eq:1soliton}). Henceforth these parameters will be
referred as phase shifts, since they are directly related to the
time-delay effect of the scattering process.

Now that we have defined the arbitrary phase shifts we start with
the soliton-soliton solution $u_{ss} $ which comes from the
two-soliton solution $u_{2s}$ when  we choose $\delta_2 = -1 /
\delta_1 $. We consider the following transformation
 \begin{eqnarray}\label{eq:equiv}
 a_2 \rightarrow - a_2 \ \ .
 \end{eqnarray}
Although this is enough to change a single soliton to an
antisoliton, this is not the case for the two-soliton solution.
The phase shifts have also to be fixed in a specific way to
complete the mapping between $u_{ss}$ and $u_{sa}$
 \begin{eqnarray}
 J_1 &\rightarrow& - J_1  \\
 K_2 &\rightarrow& 1/K_2 \ \ .
\end{eqnarray}
This effectively changes the sign of $\delta_2$ in the expression
$u_{ss}$ converting one of the solitons to an antisoliton. In
contrast with the single soliton where the antisoliton can not be
properly defined, in the two-soliton case there is a point of
reference. A distinction between a soliton and an antisoliton can
only be realised as a specific choice of the relative sign between
the parameters $a_1,a_2$ and $V$ which does not in any case lead
to topologically distinct solutions.

The same mapping between $u_{ss}$ and $u_{sa}$ can also be
achieved by making the following transformation
 \be
 a_2 \rightarrow a_2 + \pi \ \ ,
 \ee
which effectively changes the sign of all trigonometric functions
involving the parameter $a_2$ sending the solution $u_{ss}
\rightarrow -u_{sa}$ . This transformation will be used again on a
later section when we come to consider soliton reflections, to
demonstrate exactly the equivalence of the two sets of solutions.

\subsection{Breather solutions.}\label{sec:e1}
\
There are conflicting views in the literature concerning the
existence of breathers \cite{Getmanov:1977,Dorey:1995mg}. The
problem arises because the transformation $ V \rightarrow i V$
which is usually used to generate breathers from a two-soliton
solution traveling with equal and opposite velocities, does not
necessarily lead to a solution of the equations of motion. While
the technique has been widely used before on other models, the
fact that the CSG equation involves both $u$ and $\us$, implies
that naively analytically continued solutions do not necessarily
satisfy the equation of motion.

So it is not clear, for instance, that all the breather like
"solutions" of \cite{Park:1995gc} do satisfy the CSG equations of
motion. However, since the sine-Gordon is embedded in CSG model by
taking $u$ to be chargeless, the sine-Gordon breather solutions do
satisfy the CSG equations of motion. In fact a family of charged,
complex breather solutions does exist in CSG model. Although it is
quite hard to actually check if a general breather solution
satisfies the equation of motion, a trick can be used instead. We
consider the two-soliton solution of (\ref{eq:u2s}) and we demand
that this solution is even in $V$ so that is effectively a
function of $V^2$. Now the transformation $V \rightarrow i V$,
doesn't change the reality properties of the solution but simply
introduces an overall minus sign into the arbitrary parameter
$V^2$, which is irrelevant. Making the solution even in $V$, means
that a few restrictions have to be imposed. Firstly, the charge
parameters have to be taken equal or opposite according whether
$\delta_1 \delta_2$ is plus or minus one respectively. Secondly,
some of the arbitrary position parameters, have now to be fixed.
However, up until now all the arbitrary phase shifts that appeared
were either complex ($J_i$) or real ($K_i$) and there was no
distinction between the shifts that originated from the space or
time part of the phase. However, when constructing a breather
solution, by analytical continuation of the $V$ parameter a
separation between the space and time shifts is induced. All
shifts that associated with space end up as real parameters, while
time shifts become imaginary. We can restrict ourselves to shifts
only in the $x$ direction. One could also consider more general
phases which are complex and also depend on time and the parameter
V. These however correspond to either $U(1)$ rotations or time
translations which make their use obsolete. The arbitrary shift
parameters are now both real
\begin{eqnarray}
K_s = \exp \left(2 \sqrt{\b} \cos{a_s} \frac{x_s}{\sqrt{1+V^2}}
\right) & \ \nonumber
\\ J_s = \exp \left(2  \sqrt{\beta} \sin{a_s}
\frac{V y_s}{\sqrt{1+V^2}}\right)&;\ \ \ s
 = 1 ,2 \ .
\end{eqnarray}
and should be compared with the general form of (\ref{eq:shifts}).
In order to make a breather solution from the soliton-soliton case
the following relations are required
 \begin{eqnarray}
 K_1 = \pm \frac{1}{K_2} &\mbox{and}& J_1 = \mp \frac{1}{J_2} \ \
 ,
 \end{eqnarray}
where the signs in these equation are correlated.

It should be noted that more breather solutions may exist. It is
possible that through certain restrictions a more general breather
solution can be obtained, but a direct confirmation through the
equations of motion is rather difficult.

\subsection{Collapse of a Breather.}
An analysis of the quantum CSG model \cite{Dorey:1995mg} suggests
that the soliton can be identified with the elementary particle
since the vacuum of the theory and the one-soliton are not
topologically distinct solutions. Evidence of this conjecture
exists even in the classical picture. From our experience with the
sine-Gordon model, we would expect to identify the particle with
the lowest energy breather solution. It would seem to follow that
the breather whose energy and charge correspond to that of a
single particle should be equivalent to a single soliton. This
remarkable fact can be shown as follows.

We consider the static single-soliton solution
 \be \label{eq:staticsoliton}
 u_s = \frac{\cos(a) \exp (i m \sin (a) t) }{\cosh(m \cos(a) (x
 +x_0)} \ \ ,
 \ee
where $m = 2  \sqrt{\b}$. The mass of the static soliton is given
by (\ref{eq:bulkenergy}) which after integration gives
 \be
 M_{s} = {4 m \cos(a_s) }\ ,
 \ee
while the charge of the soliton is given by
 \be
 Q_{s} = i \int \frac{ \us \dt u - u \dt \us}{ 1- u \us} =
 4 \left(\mbox{sign}[a_s]
 \frac{\pi}{2} - a_s \right) \ .
 \ee
The mass of a breather is twice the mass of a single soliton
solution at velocity $V$, which has been analytically continued $V
\rightarrow i V$
\be
 M_{B} = \frac{8 m \cos (a_B) }{\sqrt{1+ {V_B}^2}} \ .
 \ee
The breather solution is effectively constructed from two
one-soliton solutions, each with charge
 \be
 Q_B =4 \left(\mbox{sign}[a_B] \frac{\pi}{2} - a_B \right) \ .
 \ee
In order to have a chance of identifying the breather with the
soliton, we demand that the mass of a breather is equal to the
mass of a static single soliton and that their charges also
coincide
\be
 M_s = M_B \ \ , \ \ Q_s = 2 Q_B \ \ .
 \ee
From the above relations, one can solve for the parameter $V_B$
 \be
 V_B = \sqrt{\frac{ 2 \cos (a_B)}{\cos (a_s)} - 1} \ .
 \ee
If this value is substituted into the breather, then the solution
collapses to a static single-soliton carrying double the charge
$Q_B$. In other words, the single-soliton can always be considered
as a bound state of two single-solitons carrying half the charge.
The argument can be used recursively so that a soliton can be
regarded as an infinite collection of solitons carrying fractions
of the original charge. At each level a soliton is identified with
a breather emerging out of a soliton pair of half the original
charge. In the classical picture this process can be carried out
indefinitely, but in the quantum case the finite character of the
mass states restricts this procedure.

This is not surprising since the static single-soliton of
(\ref{eq:staticsoliton}) can be viewed as a bound state due to the
oscillation effect which  creates a breather-like behaviour. This
is consistent with the fact that any breather can collapse to to
this solution when the parameter $V$ is properly fixed. It can
therefore be realised as a breather solution after the collapse,
exhibiting all of its former properties.

One point that has to be emphasized is that breathers constructed
with the method described in the previous section are not
chargeless. This is due to the fact that the choice of the charge
parameters $a_i$ is such that both solitons that are combined to
create a breather have the same charge. This is confirmed by the
above demonstration in which a breather collapses to a single
soliton solution which carries double the charge of the breather's
solitons. Neutral breathers do exist but only at the chargeless
limit and  can be identified with the breathers of the simple
sine-Gordon theory.

\section{Reflections.}
\
In the following sections we consider the effect of introducing a
boundary into the CSG model. Although we are free to add any
boundary potential to the Lagrangian, we choose to investigate
those potentials and their corresponding boundary conditions which
preserve integrability. Such choices allow the non-perturbative
solution of the model which is one of the main motivation for its
study. Once such suitable conditions are introduced we examine the
scattering of particles and solitons, and determine the necessary
conditions for the existence of boundary bound states.

\subsection{Abelianisation of the Lax pair and conserved currents. }
\

We shall consider a boundary condition to have preserved the
integrability of the CSG model, if we can still construct an
infinite number of commuting conserved charges. In contrast with
the theory in the bulk, the introduction of a boundary destroys
the translation invariance of the model but preserves the time
translation invariance. It is thus expected that the momentum will
not be conserved, whilst the energy will. This situation also
holds for the higher-spin conserved quantities. All energy-like,
parity-even quantities can be conserved, unlike their
momentum-like, parity-odd partners. Nevertheless, since there is
an infinite number of conservation laws, the main goal would be to
concentrate on the conservation of the parity-even quantities.

The presence of the spectral parameter $\lambda$ in the Lax pair
of (\ref{eq:lax2}) implies the existence the infinite conserved
currents in the bulk that can be determined through the method
used by Turok and Olive \cite{Olive:1985mb}. This is achieved by
performing a gauge transformation $U$
 \be\label{eq:gauge2}
  \a = U A U^{-1} +
 \d U U^{-1} \ ,
 \ee
in such a way that the commutator of the transformed gauge fields
$\a$ and $\ab$ of the Lax pair to be zero. The equation of motion
becomes
 \be
 \dt (\bar{\a} - \a)= \dx (\ab+\a) \ \ ,
  \ee
where the normal time and space derivatives are used. In the
theory in the bulk we integrate over $x$. If
$Q(\l)=\int_{\infty}^{+\infty} (\bar{\a} - \a ) dx$,  then
  \be
\frac{d}{dt} Q(\l) =  \int^{\infty}_{-\infty}  \dt(\bar{\a} - \a)
\ dx = \int^{\infty}_{-\infty}
  \dx (\ab+\a) \ dx = \left[(\ab+\a)\right]^{\infty}_{-\infty} \ \ .
 \ee
Since at infinity the fields are taken to vanish so that $\a$ and
$\bar{\a}$ approach a fixed value, it follows that the right-hand
side vanishes. As $\a$ and $\ab$ can be expanded as an infinite
Laurent series in $\l$, the coefficients of each power of $\l$,
provides us with an infinite number of conserved charges.

When a boundary is introduced the left-hand side involving the
spatial derivative does not vanish since now the integration takes
place over the semi-infinite interval. Instead one is left with an
equation of the form

 \be
 \int^{0}_{-\infty}  \dt(\bar{A} - A) \ dx = \left[(\A+A)\right]_{x=0}
 \ee
where the left-hand side is evaluated at the boundary. Instead of
demanding that the right-hand side vanishes, we instead ask that
it  can be expressed as a total time derivative with the help of
suitable conditions, thus leading to a conserved quantity.

We begin by finding explicit expressions for  ``low-spin''
conserved charges of the CSG model in the bulk by solving for the
abelianizing gauge transformation $U$ order by order in the
spectral parameter.

Let $U$ be a general real $SU(2)$ matrix, with $\det(U)=1$. The
diagonal elements of $U$ can be taken equal due to residual gauge
freedom which leave $\a$ and $\bar{\a}$ in an abelian form. Thus
$U$ takes the form
 \be
 U= \frac{1}{\sqrt{1-\x \bar{\x}}} \left( \begin{array}{cc}
   1 &  \x \\
   \bar \x & 1 \
  \end{array} \right)\ \ ,
  \ee
where $\x$ is a function of the fields and should not be
associated with the space variable. We demand that $U$
diagonalises both $A$ and $\A$ at the same time. The transformed
fields lie both in the $\sigma_3$ direction and the non-zero
diagonal elements can be identified with the conserved currents.
Taking $A$ to be
 \be
A=\left(\begin{array}{cc}
  i \Lambda & E \\
  -\Es & -i\Lambda
\end{array}\right),
 \ee
with $\Lambda=\b \l$ and $E = i ( \us \d\vs - \vs \d \us)$, we
demand that the non-diagonal part of $\alpha$ vanishes
 \begin{eqnarray}\label{eq:offdiag1}
 2i \Lambda \x +\x ^2 \Es + E +\d \x =0 \ \ ,\nonumber \\ \\
 2i \Lambda \bar{\x} + {\bar{\x}}^2 \Es + E +\d \bar{\x} =0 \ \ . \nonumber
 \end{eqnarray}
The conserved quantities can also be written in terms of $\x$ and
$\bar{\x}$
 \be
 J= -i \Lambda \frac{ 1 - \x \bar{\x}}{1 - \x \bar{\x}} - \frac{\x \Es
 +\bar{\x} E}{1 - \x \bar{\x}}+\frac{\x\d \bar{\x} - \bar{\x} \d \x}
 {2(1 - \x
 \bar{\x})} \ \ .
 \ee
The same matrix $U$, should also diagonalise $\bar{A}$, which is
given by (\ref{eq:A})
 \be
 \bar{A} = \left( \begin{array}{cc}
   D & P \\
   \Ps & -D \
 \end{array} \right) \ \ ,
 \ee
where $D = u \us - v \vs $  and $ P = -2 i \us \vs$. The choice of
$E$, $P$ and $D$ is not accidental. They actually represent the
electric field, the polarization and the population inversion
field variables respectively, when this theory is used to describe
the propagation of optical pulses in a non-linear medium
\cite{Park:1995ya}. When $U$ acts on $\A$, we again demand the off
diagonal parts to vanish. Examining the matrix explicitly yields
 \begin{eqnarray} \label{eq:offdiag2}
\frac{i}{\l}( - 2 D \x + P - \Ps \x^2)+ \db \x &=& 0 \ , \nonumber
\\ \\
 \frac{i}{\l}(2 D \bar{\x} + \Ps - P \bar{\x}^2) + \db \bar{\x} &=& 0 \
 .
 \nonumber
 \end{eqnarray}
It is easy to see that these equations are equivalent to equations
(\ref{eq:offdiag1}).

The diagonal part yields the other component of the conserved
current
\be
\bar{J} = \frac{i}{2 \l} \left( 2 \frac{1+\x \bar{\x}}{1-\x
\bar{\x}}D +2 \frac{1\Ps \x - \bar{\x} P}{1-\x \bar{\x}} \right) +
\frac{( \x \d \bar{\x} - \bar{\x} \db \x)}{2(1-\x \bar{\x})}.
 \ee
In order to solve the two sets of equations (\ref{eq:offdiag1}) or
equivalently  (\ref{eq:offdiag2}), we consider an expansion of
$\x$ and $\bar{\x}$ in powers of $\Lambda$
\begin{eqnarray}\label{eq:xseries}
\x =
\frac{\x_1}{\Lambda}+\frac{\x_2}{\Lambda^{2}}+\frac{\x_3}{\Lambda^{3}}+...
\ \ , \nonumber \\ \\\bar{\x}
=\frac{\bar{\x}_1}{\Lambda}+\frac{\bar{\x}_2}{\Lambda^{2}}
+\frac{\bar{\x}_3}{\Lambda^{3}}+...
\ \ . \nonumber
\end{eqnarray}
The coefficients $\x_i$ and $\bar{\x}_i$ can be determined by
direct substitution into (\ref{eq:offdiag1}) and
(\ref{eq:offdiag2}), and by demanding that the coefficients in all
powers of $\Lambda$ vanish. Up to order O($\Lambda^{-2}$) one
finds
\begin{eqnarray}
\begin{array}{l} \x = \left( \frac{i}{2 \Lambda} \right) E +
\left (\frac{i}{2
\Lambda} \right ) ^2 \partial{E}+ \left( \frac{i}{2 \Lambda}
\right)^3 ( E^2 E^* + \partial^2 E)+ ...  \ \ , \\ \bar{\x} =
\left( \frac{i}{2 \Lambda} \right)E^* +\left( \frac{i}{2 \Lambda}
\right)^2 \partial{E^*}+ \left( \frac{i}{2 \Lambda} \right)^3 ( E
E^{*2} + \partial^2 E^*) + ... \ \ , \end{array}
\end{eqnarray}
Now that $\x$ and $\bar{\x}$ have been defined, we can also
express the conserved quantities as a series in $\l$. Each order
of $\l$, provides a conserved quantity and since the series of
$\l$ in $\x$ and $\bar{\x}$ does not terminate, we thus have an
infinite number of conserved quantities as expected from the
integrability of the CSG model. The two components of the
conserved current up to ${\cal{O}}(\l^{-2}$) can be read off as
coefficients in the following expansion of $J$ and $\bar{J}$
\begin{eqnarray} \label{eq:Currents1}
J &=& - \l \b -\frac{i}{2 \b} E \Es \left(\frac{1}{\l}\right) -
\frac{1}{8 \b^2} (E \d \Es - \Es \d E)\left(\frac{1}{\l^2}
\right)+... \ \ ,
 \\
\bar{J} &=& i D \left(\frac{1}{\l}\right)  + \frac{1}{4 \b } ( \Es
P - E \Ps ) \left(\frac{1}{\l^2}\right)+... \ \ ,
\end{eqnarray}
and it can be checked that this current is conserved explicitly
from the equation of motion.

In the above we have constructed conserved currents that lead to
conserved charges in the bulk. However, as we have previously
argued, conserved charges on the half line are expected to take
the form of an integral over a {\it parity-even} conserved
current. The conserved currents above are neither parity even or
odd. To rectify this we note that our system of equations and
constraints possess a $Z_2$ invariance involving parity
transformations which can be used to  construct a ``reflected" set
of conserved currents. The ``reflected" set of conserved currents
is easily obtained through the substitution $\d \rightarrow \db$
in the expressions (\ref{eq:Currents1}) including those
derivatives involved in the definition of $E$. The new set of
currents $\tj$, $\tjb$, can now be combined with the former set to
produce pure parity odd and even currents.

In the presence of a boundary only parity even quantities are
conserved. The desired form of the equations to emerge is
 \be
 \dt \ \ (parity \ \ even) = \dx \ \ (parity \ \ odd).
 \ee
By combining the two sets of currents one can separate the odd and
even quantities for all powers of $\lambda$.
 \be
 \dt \left[ (\J + \tjb) -(J + \tj ) \right]=
 \dx \left[ (J - \tj) + (\J - \tjb ) \right]
 \ee
 We examine the  $\l^{-1}$ term in the expansion which gives
 \be
 \dt \left( E \Es + \tilde{E}\tilde{\Es}+ 2 \b (D+
 \tilde{D})\right)= \dx \left( \tilde{E}\tilde{\Es}- E \Es + 2 \b
 (D-
 \tilde{D}) \right),
 \ee
where $\tilde{E} = E( \d \rightarrow \db)$, etc. After integration
over the semi-infinite interval, the right hand side representing
the parity odd part is
  \be
 {\dt}{\cal W} (u, \us) =  \left(\frac{{2} \dx \us }
 { 1 - u \us}\right) \dt u  + \left(\frac{{2} \dx u }{
1 - u \us} \right) \dt \us \ \ .
 \ee
This is a total derivative provided that
 \be
 \frac{{2} \dx \us }{ 1 - u \us}=\frac{\d {\cal W} }{\d u} \ \ , \
 \ \frac{{2} \dx u }{ 1 - u \us}=\frac{\d {\cal W} }{\d \us} \ \ .
 \ee
The conserved quantity at hand, in terms of $u$ and $\us$, is then
 \be \label{eq:conserv1}
 {\cal H} = \int_{-\infty}^{0} \left({2} \frac{|\dt u|^2 + |\dx u|^2 }
 { 1 - |u|^2} + 4 \b (2 |u|^2 - 1 )
 \right) dx - [{\cal W}]_{x=0} \ \ ,
 \ee
Since this quantity actually represents the energy of the system,
${\cal W}$ can be identified with the energy contribution of the
boundary term.

When constructing the odd and even quantities of the $\l^{-2}$
term, one ends up with
 \begin{eqnarray} \label{eq:2ndcurrent}
 \dt \left( \frac{1}{2} (\E \d E - E \d \E +
 \te  \db \tes - \tes \db \te ) - \b (\E P - E \Ps + \tes \tp - \te \tps
 )\right)=  \nonumber \\
\dx \left( \frac{1}{2} ( -\E \d E + E \d \E +  \te  \db \tes -
\tes \db \te ) - \b (\E P - E \Ps - \tes \tp + \te \tps
 )\right)
 \end{eqnarray}
Once more the parity-odd right hand side which after integration
yields
 \begin{eqnarray}\label{eq:2ndrhs}
 4 \frac{\dt u \dt\dx \us }{1- u \us} -4\frac{\dt \us \dt\dx u }{1- u \us}
 - 4\frac{(\dx u \dx \us+ \dt u \dt \us)(u \dx \us - \us \dx u)}
 {(1-u \us)^2}
 \nonumber \\ -4\b (u\dx \us - \us \dx u) +4\frac{\dx u \dt^2 \us}
 {1 - u \us}
 - 4\frac{\dx \us \dt^2 u}{1 - u \us} \ \ ,
  \end{eqnarray}
should be written as a total time derivative in order to force the
currents to be conserved at the boundary. A set of boundary
conditions have to be introduced to ensure that his is the case
(App. \ref{App1}). Using the equations of motion, the parity-odd
part of (\ref{eq:2ndrhs}) can be written as a total derivative if
the following restrictions are enforced
 \begin{eqnarray} \label{eq:bc}
 \dx u =- C u \sqrt{1 - u \us } \ , \nonumber \\
 \\
 \dx \us =- C \us \sqrt{1 - u \us } \ . \nonumber
 \end{eqnarray}
The boundary constant $C$ is defined by the theory and is
responsible for the way fields react to the boundary. Consistency
of the two equations in (\ref{eq:bc}) implies that $C$ should be
considered a real parameter. When one makes the transformation
described in (\ref{eq:u2phi}), the new boundary conditions for the
fields $\p$ and $\e$ are
 \be
 \dx \p =-  C \sin(\p) \ \ \ \ , \ \ \ \dx \e = 0 \ \ ,
 \ee
which clearly shows that $C$ has to be real.

It has to be pointed out that (\ref{eq:bc}) is not the only set of
boundary conditions that can be derived. A number of isolated
``Dirichlet"-like conditions also exists. However, we restrict
ourselves only to cases where the space derivatives of the fields
appear. If we take the field $u$ to be real, the system is reduced
to the sine-Gordon equation with a boundary condition $\dx \p = -
C \sin \p$ . This is the subset of integrable boundary conditions
of the sine-Gordon model presenting the $Z_2$ symmetry $ \p
\rightarrow -\p$. The corresponding conserved quantity, is rather
large and is omitted.

\section{Soliton scattering and boundary bound states}
Since the necessary conditions for the integrability of the model
have been established, we study the scattering of particles and
solitons off the boundary. We begin this section with the effects
of introducing a boundary potential to the vacuum of the theory.
We continue with the scattering of particles and solitons and
derive the phase shifts induced by the process. Finally, we
investigate the necessary conditions for the existence of boundary
bound states.

\subsection{Vacuum}
When a boundary term is introduced, the vacuum of the theory that
we discussed in section (2.2), does not necessarily remain
unchanged. It is exactly this contribution that needs to be
carefully examined before any statements are made about the
minimum energy configuration. Although, the vacuum solution of the
theory in the bulk is a strong candidate, soliton solutions could
also be considered in the attempt to both minimize the energy
functional and satisfy the boundary conditions of (\ref{eq:bc}).

We begin by first determining the energy contribution of the
boundary term. The full Lagrangian of the model is now
 \be
 {\cal L}_{tot} =  {\cal L} + \cal{ L}_B .
 \ee
The boundary term $\cal{ L}_B $, can be determined by the
variation principle of the total action. The variation of the
${\cal L}$ term yields
 \be
{\cal \delta L} = \frac{ \d {\cal  L} }{\d u} \delta u +\frac{ \d
{\cal  L} }{\d \us} \delta \us + \frac{ \d {\cal  L} }{\d
(\d_{\mu}u)} \delta (\d_{\mu} u) + \frac{ \d {\cal  L} }{\d
(\d_{\mu}\us)} \delta (\d_{\mu}\us) \ .
 \ee
When the Euler-Lagrange equations are used, two terms survive
since the model is considered in the semi-infinite interval where
the fields do not vanish at the boundary
 \be
 {\cal \delta L} = \d_{\mu} \left(\frac{ \d {\cal  L} }{\d
 (\d_{\mu}u)} \delta  u \right) + \d_{\mu} \left(\frac{ \d {\cal  L} }{\d
 (\d_{\mu}\us)} \delta  \us \right) \ \ .
 \ee
From the variation of the boundary term one has
 \be
 {\cal \delta { L}_B} = \frac{\d {\cal{ L}_B} }{ \d u} \delta u
 +\frac{\d {\cal{ L}_B} }{ \d \us} \delta
 \us \ .
 \ee
The variation of the action vanishes when the remaining terms
evaluated at the boundary are forced to cancel. The two
interrelated equations that emerge are
 \begin{eqnarray}
 \frac{ \d {\cal  L} }{\d
 (\dx \us)} = \frac{ - \dx u}{1-u \us}  = - \frac{\d {\cal {
 L}_B}}{\d \us} \ \ , \\
\frac{ \d {\cal  L} }{\d
 (\dx \us)} = \frac {- \dx \us}{1-u \us}  = - \frac{\d {\cal {
 L}_B}}{\d \us} \ \ .
 \end{eqnarray}

By substituting the boundary conditions of (\ref{eq:bc}), these
can easily be solved for the boundary term
 \be
 {\cal L}_B = 2 C \sqrt{1 - u \us} \ .
 \ee
We now consider the total energy of the system , now comprising of
two parts
 \be \label{eq:totenergy}
 {\cal H}_{tot} = {\cal H}_{bulk} + {\cal H}_B \ ,
 \ee
where the term ${\cal H}_{bulk}$, represents the energy in the
bulk and  the second term ${\cal H}_B$ represents the  energy
contribution from the boundary
 \be \label{eq:HB}
 {\cal H}_B = - 2 C \sqrt{ 1 - u \us } \ \ ,
  \ee
which is evaluated at $x=0$. This energy contribution makes the
determination of the vacuum difficult. The sign of the boundary
constant $C$ is not set, which could provide either a positive or
negative contribution to the total energy of the system. This
clearly shows that although the original choice for a vacuum
should not be discarded, one should also consider other static
solutions which in conjunction with the sign of $C$ could provide
a lower energy vacuum than before.

Apart from the original choice for a vacuum, one can consider
static multi-soliton solutions. We restrict ourselves to
one-soliton solutions since experience with similar models usually
makes multi-soliton solutions unsuitable candidates.

When considering one-soliton solutions, one has the equations of
the \B transformation (\ref{eq:back1}) which are always true to
simplify expressions. In particular we first consider the ${\cal
H}_{bulk}$ term representing the energy in the bulk
 \be
 {\cal H}_{bulk} = \int dx \left(\frac{|\dt u|^2 +|\dx u|^2}{1 - u \us}
 + m^2 u \us \right) \ \ ,
 \ee
with $m=2 \sqrt{\b}$. By direct substitution of the \B equations a
simplified expression of the bulk energy is acquired
 \be \label{eq:bulkenergy2}
 {\cal H}_{bulk} = \int dx ( 2 m^2 u \us) \ .
 \ee
When the above expression is integrated throughout space, the
result can be identified with the mass of the soliton solution
$u$. However, now the integration is over the half line and
specifically over the $[-\infty,0]$ region.

The same equations can be used to express the ${\cal H}_{B}$ term
of (\ref{eq:HB}) . Specifically, the boundary constant $C$ is
determined by direct comparison of the \B equations of
(\ref{eq:back1}) and the boundary condition which appears in
(\ref{eq:bc})
 \be
 C =  \frac{m}{2} \left( \delta \ e^{i(\t + \Omega)}
 +\frac{1}{\delta} \  e^{-i(\t + \Omega)} \right) \ .
 \ee
with $\t$ given by (\ref{eq:theta}). At $x=0$ and assuming that
$V=0$, the above expression simplifies to
 \be \label{eq:C}
 C = \pm \frac{m}{\sqrt{1 + \tan^2(a) \coth^2 (m \cos (a) x_0)}} \ .
 \ee
This implies that $|C|\leq |m|$. Since both $m$ and $C$ are
defined by the theory, the above relation is true only for
specific choices of the boundary parameter $C$. Alternatively, one
can think of this restriction emerging from the fact that for $|C|
> |m| $, no choice of $x_0$ satisfies the boundary condition.

In the case where we choose $u=0$ as a possible vacuum, the only
remaining term in the total energy is
 \be
 {\cal H}_{tot} = - 2 C \ .
 \ee
Alternatively, one can consider a one-soliton solution where V is
set to zero which appears in (\ref{eq:staticsoliton}). In this
case, both terms of (\ref{eq:totenergy}) depend on the initial
position of the soliton. However, after some calculations, the
$x_0$ dependence drops out and the total energy is given by a the
following expression
 \be
 { \cal H}_{tot} = 2 m \cos (a) \ .
 \ee
It is far from obvious, which vacuum choice provides the minimum
energy configuration. To determine this, one has to look at the
expression of the boundary constant $C$ in (\ref{eq:C}). This can
be rewritten in the following form
 \be \label{eq:vfun}
 y^2+ \frac{ y^2 -y^4}{y^2 +F^2} = \cos^2(a) \ ,
 \ee
where
 \be
 F^2 = \sinh^2 (m \cos (a) x_0) \ \ , \ \ y = \frac{C} {m} \ .
 \ee
In the above relation $m$ and $C$ should be treated as fixed
parameters, while $F$ can be varied through $x_0$. The left hand
side of (\ref{eq:vfun}) is monotonically decreasing as  $F$
increases since $0< y^2 < 1$. We observe the following
\begin{eqnarray}
\cos^2(a) = 1  &\mbox{when}& \ \ F \rightarrow 0 \ \ ,\nonumber \\
& &
\\ \ \cos^2(a) = y^2 &\mbox{when}& \ \ F \rightarrow \pm \infty \
\ . \nonumber
\end{eqnarray}
This shows that moving the soliton away from the boundary
decreases the energy of the system. On the extreme case where the
soliton is placed at infinity, the model behaves as if no soliton
exists, and the only contribution is the boundary term which
coincides with the vacuum solution of $u=0$. On the contrary as
the soliton is placed closer to the boundary the energy increases.
The maximum energy occurs when $F=0$ at which point $\cos^2(a) =1$
so that $H_{tot} = 2 m$ which is greater than the energy
$H_{tot}=2C $ of the $u=0$ vacuum.

Although the choice of vacuum in the bulk seems to be the most
suitable choice in the boundary case too, one cannot rule out
multi-soliton solutions that might provide lower energy
configurations. This demands tedious calculations and remains as
one of the open questions for this model.

\subsection{Soliton reflections}
In this section we investigate the reflection of solitons from the
boundary. Mathematically this can be represented by a two-soliton
solution satisfying the boundary condition. One of the solitons
represents the incoming soliton whilst the other represents the
reflected one. The point where the two solitons actually meet
along the whole line as well as the phase shift due to their
collision create an overall time-delay effect which can be
calculated directly through the parameters of the scattering. This
time-delay can be attributed to the interaction of the soliton
with the boundary.

However, the most difficult step is to determine the restrictions
that have to be imposed so that the two-soliton solution satisfies
the boundary condition
 \be \label{eq:bc_2s}
 \dx u_{2s} = - C u_{2s} \sqrt{1- u_{2s} {u^{*}_{2s}}} \ .
 \ee

Energy and charge conservation laws demand that both the mass and
the charge of the soliton are conserved by the boundary. This
restricts the choice of the charge parameters $a_1, a_2$ to be
either equal or opposite.

Due to the large expressions involved in the calculation, one is
forced to expand both sides of the equation (\ref{eq:bc_2s}) to a
Taylor series in exponentials of $t$, and match each term of the
same order. Each term provides us with an equation involving the
boundary parameter $C$. As mentioned in the previous section, the
boundary constant has to be a real parameter. The real and
imaginary parts of the equation yield two constrains on the
parameters.

Let us consider this in more detail. We begin with a two-soliton
solution, where the parameters are chosen in such a way so as to
describe a soliton-soliton scattering. In this case, the charge
parameters are taken to be opposite $a_1=-a_2$ and the \B
parameters to be $\delta_1=-1/ \delta_2$.

Furthermore, we adopt the following parametrisation which is more
natural
 \be
 \frac{K_1}{K_2} = e^\l \ \ , \ \ \frac{J_1}{J_2} = e^{i \zeta} \ \
 ,  \ \ V = \tanh (\vt) \ \ .
 \ee
After both sides of the boundary equation are expanded as a Taylor
series in time, we can discard the imaginary parts from all terms
by using the following relation
 \be \label{eq:res1}
 \sin(\zeta)= - \tanh(\vt) \tan(a) \sinh(\l).
 \ee
When the above equation is used the infinite set of equations
collapse to a single constraint
 \be \label{eq:res2}
 C =  m \cos(a) \cosh(\vt)\ \left(\frac{ \cos(\zeta)+\cosh(\l)
 }{\sinh(\l)}\right)\ \ .
 \ee
When the shift parameters are fixed according to the above
relations, the two-soliton solution satisfies the boundary
condition and this process describes a soliton being reflected by
the boundary.

The fact that only relative shifts in both normal and internal
$U(1)$ space are important should be expected from time
translational and $U(1)$ invariance of the model. The
non-topological solitons in the CSG theory are reflected as
solitons carrying the same charge $Q$. This is because the
boundary potential does not break the $U(1)$ symmetry since it
depends only on $|u|$.

\begin{figure}[htb]
\begin{center}
\fbox{
\includegraphics[width=.75\textwidth,height=.5\textwidth]{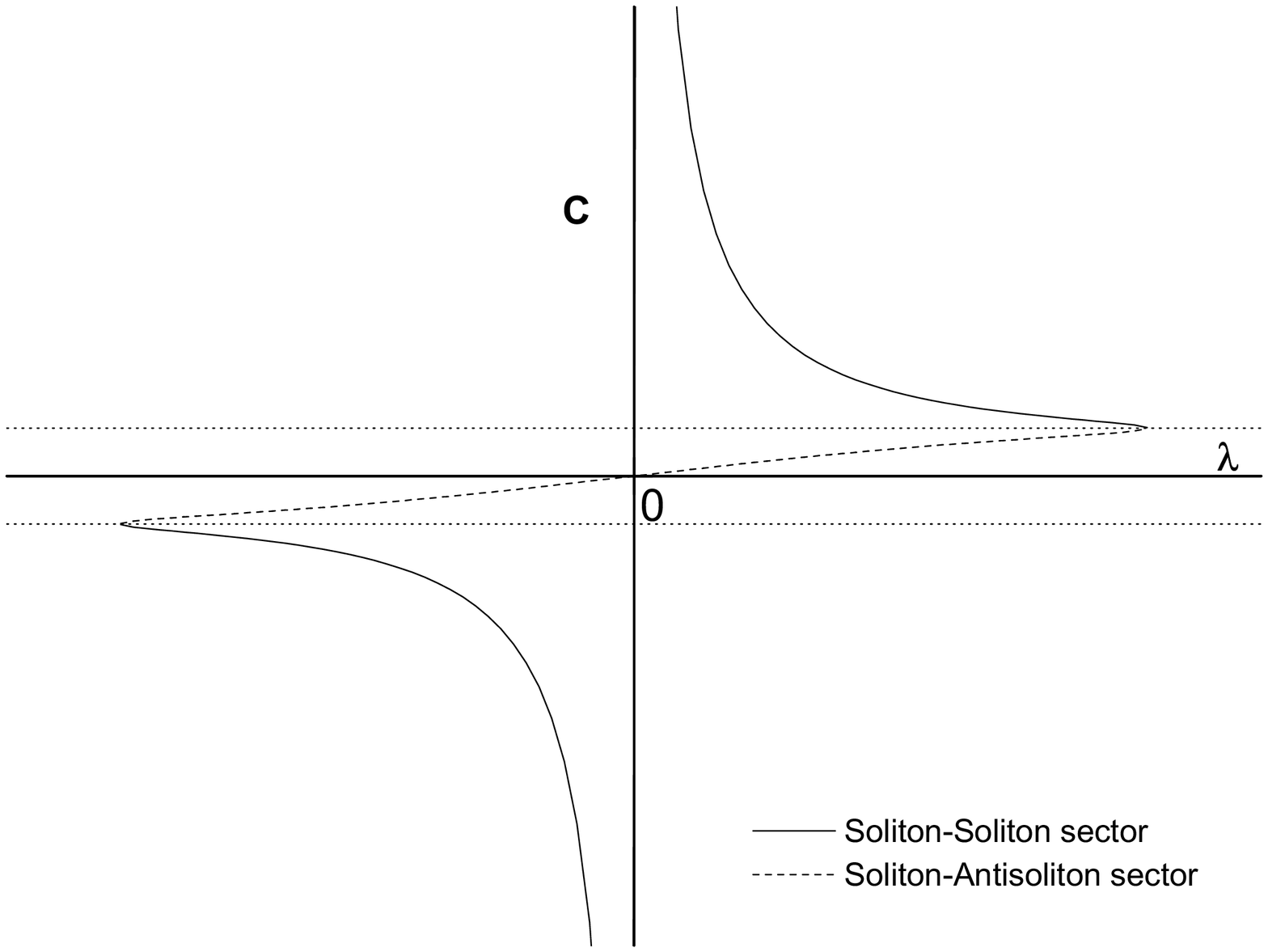}}
\end{center}
\caption{$C$ in terms of $\l$} \label{plot1}
\end{figure}

We shall consider $a$ and $\vt$ as fixed parameters and use
equations (\ref{eq:res1}) and (\ref{eq:res2}) to determine the
parameter  $\l$ in terms of $C$. By eliminating the parameter
$\zeta$, one ends up with the following quadratic equation for $C$
 \be
{C}^{2}-2\,{\frac { C m \cos(a)\cosh(\vt)\cosh(\l)}{\sinh(\l
)}}+\left (\sinh(\vt)^{2}+\cos(a)^{2} \right ){m}^{2}
 =0 \ \ .
 \ee

The solutions of the above equation can be plotted to present the
dependence on $\l$. The plot involves two branches (Fig.
\ref{plot1}) due to the sign ambiguity, which are mutually
exclusive. The plot shows that a soliton can always be reflected
by the boundary. The branches meet at the points
 \be
 C= \pm m \sqrt{\cos^2(a) + \sinh^2(\vt)} \ \ , \ \ \coth(\l) =
 \frac{\sqrt{\sinh^2(\vt) +\cos^2(a)}}{\cos(a) \cosh(\vt)}\ \ .
 \ee

\begin{figure}[htb]
\begin{center}
\fbox{
\includegraphics[width=.75\textwidth,height=.5\textwidth]{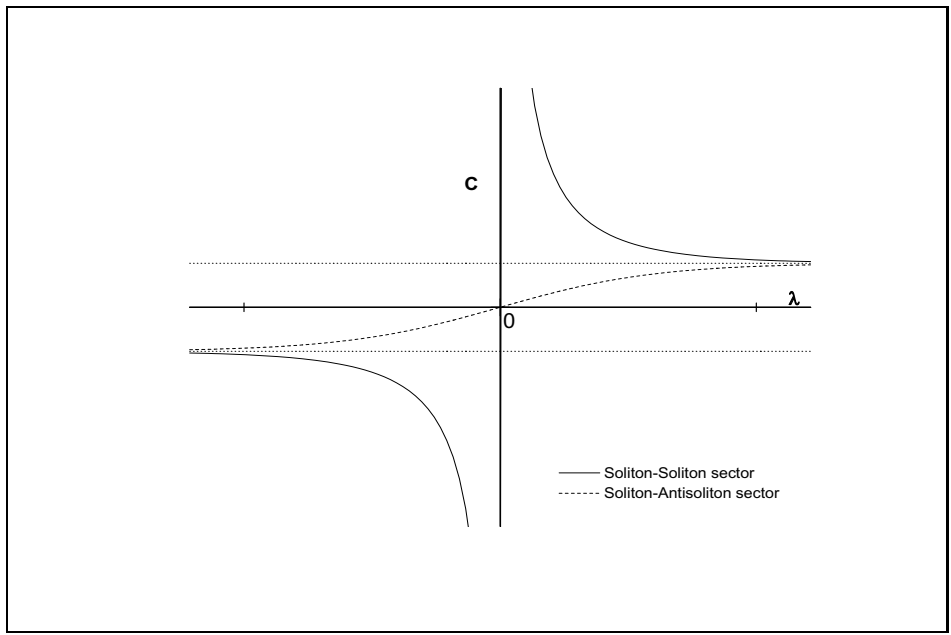}}
\end{center}
\caption{$C$ in terms of $\l$ for chargeless case} \label{plotSG}
\end{figure}

In the limit $a \rightarrow 0$ the two branches of the plot can be
identified with  the soliton-soliton and soliton-antisoliton
sector of the reflection process at the sine-Gordon limit (Fig.
\ref{plotSG}). For fixed values of $\vt$ and $a=0$, it is the
value of the boundary constant $C$ which determines whether a
soliton is reflected as a soliton or an antisoliton.  For $C$
small, a soliton is reflected as an antisoliton (Neumann boundary
conditions for $C=0$), while for $C$ large a soliton is reflected
as a soliton (Dirichlet boundary conditions for $C=\infty$). For
$C= m \cosh(\vt)$ the branches do not meet as in the CSG case.
This specific value of $C$ corresponds to a logarithmic divergence
that appears in the classic time delay for the sine-Gordon case.
These results coincide with the results derived by previous
treatments of the boundary sine-Gordon model \cite{Saleur:1995yh}.

\subsection{The classical time delay}
The time delay which appears at the scattering of a soliton off
the boundary, can be calculated directly from the asymptotic
values of the solution at $t=\pm \infty$. We begin with the
two-soliton solution and change to a frame of reference which
moves with the incoming soliton (i.e. $x = V t+\tilde{x}$. In the limit
$t=-\infty$ the solution becomes
 \be \label{eq:Sminus}
 S_{-} = \lim_{t\rightarrow - \infty} u_{2s} =
 \frac{\cos(a) \ e^{i (A_1+B_1)}}
 {\cosh( P(-\tilde{x}+x_1)+r)} \ \ ,
 \ee
where
 \be
  A_1= \frac{P \sin(a)}{\cos(a)}\left[ (1-V^2) t -V \tilde{x} - y_1\right] \ \
 , \ \ \tan(B_1)= -\frac{V \sin(a)}{\cos{a}}
 \ee
and
 \be
 P = \frac{m \cos (a) }{\sqrt{1-V^2}} \ \ , \ \
 \\ \ r = \frac{1}{2} \ln
 \frac{V^2}{\cos^2(a)+V^2 \sin^2(a)} \ \ .
 \ee
The parameters $x_i$ and $y_i$ represent regular shifts that were introduced in (\ref{eq:shifts}). The solution, as expected, describes a single incoming soliton at
early time far away from the boundary.

We repeat the same calculation, but now we change to the frame of
reference of the outgoing soliton (i.e. $x = - V t $) and
calculate the limit of the two-soliton solution at $t=+ \infty$
which yields
 \be \label{eq:Splus}
 S_{+} = \lim_{t=+\infty} u_{2s} = \frac{\cos(a) \ e^{i(A_2+
 B_2+\pi)}}{\cosh(P(-\tilde{x}-x_2)+q)} \ \ ,
 \ee
where
 \be
 A_2 = \frac{P \sin(a)}{\cos(a)}\left[ (1-V^2) t + V \tilde{x} - y_2\right] \ \
 , \ \ \tan(B_2)=\frac{V \sin(a)}{\cos(a)} \ \ ,
 \ee
and
 \be
 \ \ q = \frac{1}{2} \ln
 \frac{V^2}{\cos^2(a)+V^2 \sin^2(a)} \ \ .
 \ee
Once again this is a single soliton solution representing the
reflected soliton far away from the boundary wall.

The asymptotic solutions $S_{+}$, $S_{-}$ contain all the
information needed to calculate the time-delay. The latter is a
combination of two separate events. Firstly, a phase shift is
induced during the scattering of the two solitons. Before the two
solitons re-emerge as two separate entities, the reconfiguration
of the solution creates a phase shift which is equivalent to a
time delay. Secondly, the centre of mass of the two-soliton
solution does not necessarily lies at the boundary. This implies
that the two solitons actually meet at a different point than
$x=0$. This creates again a time delay which may be either
positive or negative corresponding to an attractive or repulsive
boundary potential respectively.

\begin{figure}[htb]
\begin{center}
\fbox{
\includegraphics[width=.75\textwidth,height=.5\textwidth]{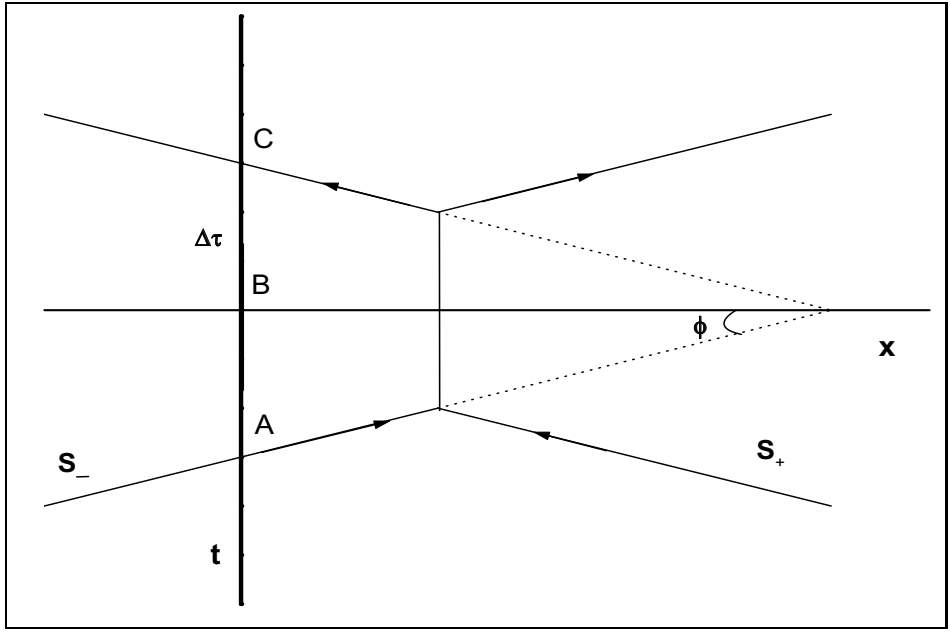}}
\end{center}
\caption{Scattering diagram}\label{fig:scat1}
\end{figure}

Ignoring any interaction between the two solitons, one can project
the trajectories of  $S_{+}$ and $S_{-}$ on $x-t$ diagram and find
the point where these cross (Fig. \ref{fig:scat1}). The distance
of this point from the boundary is proportional to the time delay
$\Delta \tau$ which in the diagram is given by the distance (AC).
The time delay corresponds to the time interval in which the
soliton appears to be absorbed by the boundary before it reemerges
as a well defined entity . The two solitons move across the
following lines
 \begin{eqnarray}
 S_{-} :\ \ \ &t& = \frac{1}{V} \left(P(-x + x_1) +r \right) \ \ , \nonumber \\
 S_{+} :\ \ \ &t& = \frac{1}{V} \left(P( x + x_2) -q \right) \ \ ,\nonumber
 \end{eqnarray}
as dictated by (\ref{eq:Splus}) and (\ref{eq:Sminus}). The lines
cross at
 \be
 x_0 = \frac{1}{2} \left( x_1 - x_2 + \frac{r+q}{P} \right ) =
 \frac{x_1-x_2}{2}+\frac{r}{P} \ \ ,
 \ee
since $r=q$. The time delay is finally
 \be
 \Delta \tau_{CSG} = \frac{ 2 x_0}{V} = \frac{(x_1-x_2)}{V}
 + \frac{\sqrt{1-V^2}}{m V \cos(a)} \ln {\left(\frac{V^2}{\cos^2(a)+V^2
 \sin^2(a)}  \right)} \ \ .
 \ee
In the expression above, the first term of the right-hand side
represents the time delay caused by the non-symmetric character of
the solution with respect to the boundary. In the special case
where $x_1=x_2$, the centre of mass lies on the boundary and the
term vanishes. The second term is independent of the initial
position of the two solitons or the boundary potential and is
caused by the phase shift of the scattering process.

The relative position of the two solitons are however fixed
according to the constraint equations  (\ref{eq:res1}) and
(\ref{eq:res2}) which ensure that solution satisfies the boundary
condition. Specifically the parameter $\l$ corresponds exactly to
the $x_1-x_2$ difference up to the overall factor $P$. It is thus
possible to express the time delay in terms of the boundary
constant, by solving the constraint equations and substituting the
relative position of the solitons. We choose to express the
velocity parameter $V$ in terms of the rapidity $\vt$ for
simplicity reasons. After a few straightforward calculations we
recover the following expression for the time delay

 \be
 \Delta \tau_{CSG} =
 \frac{\ln Q}{{2 m \cos(a) \sinh (\vt)}}
 \ee
where
 \be
 Q={\frac{\sinh^4 \vt \left((\cos^2(a) + \sinh^2 (\vt))m^2
 +2 C m \cos(a) \cosh (\vt) + C^2\right)}
 {(\cos(a)^2
 +\sinh^2(\vt))^2 \left((\cos^2(a) + \sinh^2 (\vt))m^2 -2 C m \cos(a) \cosh (\vt)
 + C^2\right)}}
 \ee
In the special limit of $a=0$, the time delay for the sine-Gordon
model is recovered
 \be
 \Delta \tau_{SG} = \frac{1}{m \sinh (\vt)}
 \ln \left({\tanh^2 (\vt) \frac{m \cosh(\vt)+C}{m
 \cosh(\vt)-C}}\right) \ \ .
 \ee
This is exactly the time delay calculated for the sine-Gordon
theory in the presence of a boundary \cite{Saleur:1995yh} for
the restricted class of boundary conditions which admit $\p=0$ as
a vacuum to which the chargeless limit of CSG correspond.

\subsection{ Boundary bound states}
In this section we examine the spectrum of bound states. Once
again, for the the boundary condition to be satisfied we need to
restrict some of the parameters in the solution.

The simplest bound state that we can have is the static single
soliton that was introduced in (\ref{eq:staticsoliton}). The
solution is not really static, as the imaginary phase survives the
setting of the speed parameter $V$ to zero. The solution is static
only in the sense that the centre of mass doesn't translate in the
$x$ direction, although the wave oscillates with fixed angular
velocity $ \omega = m \sin(a)$.

When a boundary is introduced a static soliton can satisfy the
boundary condition for $|C| \le|m|$ when its position is fixed
according to equation (\ref{eq:C}). At the chargeless limit any
time dependence vanishes and the solution collapses to a static
single soliton of the sine-Gordon theory, fixed at the boundary.

Breathers that have been constructed by the method described in
section (\ref{sec:e1}) can also be shown to satisfy the boundary
condition. The condition that $C$ is real still holds. However, all
the arbitrary phase shifts are now real numbers and constrained.
We examine breather solutions that emerge from the soliton-soliton
case. Just as before, the solution does satisfy the boundary
condition with some restrictions involving the arbitrary
parameters. Once more a Taylor expansion of the boundary equation
is needed. The parametrization used in this case is
 \be
 K_1=e^{\l} \ \ , \ \ J_1 = e^{\zeta} \ \ , \ \ V=\tan(\vt) \ \ ,
 \ee
while the parameters $K_2$ and $J_2$ have been been properly fixed
 so that this is  a breather solution. The first restriction
needed for the solution to satisfy the boundary condition is
 \be
 \sinh(2\zeta) = - \tan(a) \tan(\vt) \sinh(2\l) \ \ .
 \ee
The parameter $\vt$ plays the role of the rapidity, which has now
been analytically continued. The second restriction which
completes the necessary requirements for a boundary bound state is
 \be
 C= m \cos(a) \cos(\vt) \left(\frac{\cosh(2 \zeta)-\cosh(2\l)}{\sinh(2\l)} \right).
 \ee
Both relations can be recovered by analytical continuation of the
corresponding relations of (\ref{eq:res1}) and (\ref{eq:res2})
after the necessary restrictions for a breather solution have been
already taken into account.
\begin{figure}[htb]
\begin{center}
\fbox{
\includegraphics[width=.75\textwidth,height=.5\textwidth]{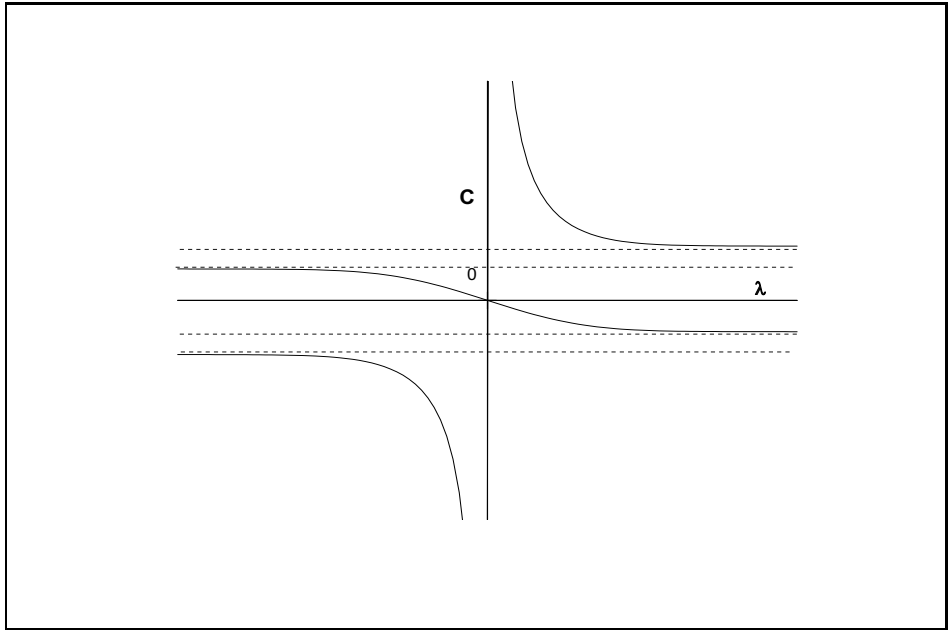}}
\end{center}
\caption{C in terms of $\l$ for Breather case}
 \label{fig:CLbreather}
\end{figure}
It is instructive to examine the relation between the parameters
$C$ and $\l$ (Fig. \ref{fig:CLbreather}) since it provides
valuable insight to the structure of bound states. There are two 
distinct regions of values of
$C$ that do not correspond to any bound state. This regions are
defined by the limit values
 \be
 C = \pm m \left(\cos (a) \cosh (\t) \pm \sin (a) \sinh (\t)
 \right) \ \ .
 \ee
At the chargeless limit, the regions collapse to the single values
 \be
 C = \pm m \cosh \t \ \ ,
 \ee
which coincided with the logarithmic divergence appearing in the
time delay for the soliton reflection.

\subsection{Particle Reflections.}
In this section we consider the spectrum of particles and their
reflection factors in the presence of a boundary.

For small fluctuations around the vacuum $u=0$, the boundary
condition (\ref{eq:bc}) becomes
 \be
 \partial_1 \epsilon(x,t) = - C \epsilon(x,t) .
 \ee
In order to calculate the reflection factor when particles bounce
of the boundary wall, we substitute in the last relation the
particle solutions presented in (\ref{eq:planewave}). The constant
$A$ of the right propagating waves is taken to be one, since it
has to do with the characteristics of the particle beam. The
reflection factor is identified with the constant $B$, which
corresponds to a phase change as the particles encounter the
boundary
 \be
 B = \frac{ i k + {\it C }}{ i k - {\it C }} \ .
 \ee
The reflection factor, as expected, depends on $C$ which as stated
before, appears as a free, real parameter in the boundary
condition. For $C = 0$, the reflection factor is equal to $B=1$
and no phase appears between the two waves upon their scattering
off the boundary. This is consistent with the fact that the
boundary term is proportional to the boundary constant, so when
$C$ is set to zero, the boundary term vanishes.

Particle solutions can be related to bound states through the pole
appearing in $B$. Indeed one may choose $k =- i C $ and apply this
to a solution of the form $\frac{1}{B} \epsilon(x,t)$. The
remaining terms depend explicitly on the boundary constant
 \be
\epsilon(x,t) =  {e^{-i\left (\sqrt {{m}^{2}-{C}^{2}}t+-iCx\right
)}} \ \ .
 \ee
This solution is square integrable only for a specific range of
values for the boundary constant. Specifically if $C$ is positive
then the solution is not square integrable since it is
exponentially increasing as $x \rightarrow \infty$.

When $-m < C < 0 $ , then $\epsilon(x,t)$ represents a square
integrable  exponentially decreasing solution as $x \rightarrow
\infty$. It oscillates with constant angular velocity
$\omega=\sqrt{m^2-C^2}$ and is therefore a stable bound state. It
can also be viewed as the tail of a static one-soliton solution
satisfying the boundary condition, with the parameters adjusted in
such a way its centre of mass goes to positive infinity. Examining
the condition (\ref{eq:C}) for the static soliton to obey the boundary
condition, this limit can be achieved as $x_0 \rightarrow \infty$,
i.e. we must take the charge is such a way that $C=-m\cos a$.

Finally in the region $C<-m$, the solution can increase
exponentially in time. This shows that the vacuum solution $u=0$
is no longer  stable. In fact the particle behaviour which
corresponds to a small perturbation around the vacuum seems to be
ever increasing. This instability can be understood through a
rather impressive mechanism in which a chargeless soliton is
emitted from the boundary, effectively changing the value of $C$
so that $u=0$ is now stable.

Recall from section (\ref{sec:sad}) that for a chargeless soliton
we should take the opposite sign for $\sqrt{1-u \us}$ on each side
of the centre of the soliton where $|u|=1$. The instability can be
viewed as a left moving chargeless soliton which approaches the
boundary from $x=\infty$. In the beginning while the soliton is
far away from the boundary $u=0$ so that the boundary potential of
(\ref{eq:HB}) is $H_B = - 2 C $. As the centre of the soliton
passes through $x=0$, the sign of the square root in the boundary
potential changes. As the soliton moves to $x=-\infty$, $u$
returns to $0$ near the boundary but now we take the boundary
energy with the opposite sign $H_B=2 C$. Effectively the sign of
$C$ has been flipped to a positive value. The energy released from
the boundary is $4C>4m$, which is greater than the rest mass of a
single chargeless soliton. At $C=-m$, the soliton is emitted with
infinitesimal velocity. As $C$ decreases, more energy is given up
by the boundary and the soliton can be emitted with larger $V$.
This process agrees with the infinite time-delay effect which was
encountered in the soliton reflections section. The soliton
emission represents the time reversal picture of that effect in
the chargeless soliton case (Fig. \ref{plotSG}). It follows that
we need never consider the situation where $C<-m$.

\section {Discussion}

The CSG model is one of the simplest generalisations of the sine-Gordon theory, but nonetheless has a rich and fascinating mathematical structure.

In the first part of the paper we have examined the spectrum of the theory in the bulk and written down explicit two-soliton solutions within the framework of the matrix potential. We also demonstrated how to construct breather solutions in an elegant way avoiding the problems that arise by the analytical continuation of the parameter $V$.
There are two ways in which soliton solutions of the CSG model differ qualitatively from those of the Sine-Gordon model. Firstly, the solitons can be charged, so that whilst there
centre of mass remains fixed, the solutions are not static. The second feature is that the CSG does not possess degenerate vacua, and so the solitons are non-topological. There is therefore no distinction between solitons and antisolitons which can be interchanged by a continuous variation of the charge parameter $a$. Nevertheless, the topological nature of the sine-Gordon theory can be recovered as the choice of branch cut of $\sqrt{1-|u|^2}$ in the chargeless limit as was demonstrated in section \ref{sec:sad} . A direct consequence of the non-topological nature of the CSG soliton is that the breather solution can collapse to a single soliton when the parameter V is properly fixed, tying in with the picture presented in \cite{Dorey:1995mg} that particle and solitons can be identified in the quantum limit.

In the second part of the paper we introduced a boundary term in the CSG Lagrangian and demanded that the system remains integrable. First we constructed low-spin conserved quantities of the theory using abelianisation of the Lax pair, and then derived suitable boundary conditions in order to preserve these. In the presence of a boundary, we examined the vacuum structure and showed that the bulk vacuum $u=0$ remained the true vacuum in the boundary case.

Soliton reflections off the boundary were also studied and the necessary constraint
equations were written down in terms of the phase shift parameters. The set of equations
was derived by demanding that the two-soliton solution satisfies the boundary condition.
Moreover the time delay induced by the scattering process was calculated in terms of the
boundary constant C and was found to coincide in the chargeless limit with the time delay
of the sine-Gordon theory.

Finally we looked for classical solutions corresponding to boundary bound states. We found that it was possible to construct both
bound soliton and bound breather solutions. We also found a bound state in the particle spectrum. This was unstable when parameter $C$ 
associated with the boundary energy was in the range $C<-m$. In this case the boundary emits a chargeless soliton, effectively
changing the sign of the boundary parameter $C$ to $C>m$.  

We end our discussion by pointing out a few aspects of the model that appear quite
interesting and deserve further study. As mentioned in the introduction the CSG theory
was used to generalise the existing field theory approach of optical pulse propagating in a
non-linear medium. A physical interpretation of the results appearing in this paper would
be extremely interesting, especially the physical meaning of breather solutions and their application to physical geometries.

An obvious extension of our results is to consider the quantum case of the boundary CSG model. The S-matrix for the model in the bulk, which corresponds to perturbed $Z_n$ parafermions, is known. It would be interesting to see if one can find a quantum reflection matrix, compatible with this S-matrix and with the classical results presented in this paper. As the simplest case in the family of homogeneous sine-Gordon theories, the results might shed light on the more complicated models in the family.

\subsection*{Acknowledgements}
We would like to thank Patrick Dorey and Roberto Tateo for illuminating discussion during the preparation of this work.
\newpage
\appendix
\section{Appendix}
\subsection{The boundary condition from  the $\l^{-2}$ term}
\label{App1}

In order to construct an infinite number of conserved quantities
when a boundary is introduced, we need to express the parity odd
terms of the equations of motion as  total time derivatives. The
corresponding term of order $\l^2$ is in terms of the fields $u$
and $\us$
\begin{eqnarray}
-2 \,{\frac {\left (\dx{\it \us}\right )\left (\dt^{2}u+
\dx^{2}u\right )}{1- u{\it \us}}}-4\,{\frac {\left (\dt {\it
\us}\right )\dx \dt u}{1-u{\it \us}}}+2\,{\frac {\left (\dx
u\right )\left (\dt^{2}{\it \us}+ \dx^{2}{\it \us}\right
)}{1-u{\it \us}}} +4\,{\frac {\left ( \dt u\right )\dx \dt {\it
\us}}{1-u{\it \us}}}  \nonumber
\\
+ 2\,{\frac {{\it \us}\,\left (\dx u \right )^{2}\dx {\it
\us}}{\left (1-u{\it \us}\right )^{2}}}-2\,{\frac {u\left (\dx
{\it \us}\right )^{2}\dx u}{\left (1-u{\it \us}\right )^{2}}}
 -4 \,{\frac {u\left (\dt {\it \us}\right )\left (\dx
{\it \us}\right )\dt u}{\left (1-u{\it \us}\right )^{2}}}
  +2\,{\frac {{\it \us}\,\left (\dt u\right )^{2} \dx
{\it \us}}{ \left (1-u{\it \us}\right )^{2}}}
 \nonumber
\\-2\,{\frac {u\left
(\dt {\it \us}\right )^{2}\dx u}{\left (1-u{\it \us}\right
)^{2}}}+4\,{ \frac {{\it \us}\,\left (\dt u\right )\left (\dx u
\right )\dt{\it \us}}{\left (1-u{\it \us}\right )^{2}}}+4\,\left
(u\dx{\it \us}-{\it \us}\,\dx u\right )\beta \nonumber \ \ .
\end{eqnarray}

Since we need the above expression to be a total time derivative
we can eliminate any second order spatial derivatives of the
fields by using the equations of motion of (\ref{eq:CSGeq1u})
\begin{eqnarray}\label{eq:2ndrhs2}
 4 \frac{\dt u \dt\dx \us }{1- u \us} -4\frac{\dt \us \dt\dx u }{1- u \us}
 - 4\frac{(\dx u \dx \us+ \dt u \dt \us)(u \dx \us - \us \dx u)}{(1-u \us)^2}
 \nonumber \\ -4\b (u\dx \us - \us \dx u) +4\frac{\dx u \dt^2 \us}{1 - u \us}
 - 4\frac{\dx \us \dt^2 u}{1 - u \us} \ \ ,
\end{eqnarray}
We take advantage of the fact that we are free to add total time
derivatives on this expression, since this represents a
conserved quantity. The expression simplifies significantly by
adding the following term
 \be
  \dt  \left(4 \frac{\dx \us \dt u - \dx u \dt \us}{1- u \us}
  \right) \ \ ,
 \ee
which yields
\begin{eqnarray} \label{ap:l21}
 -4\,{\frac {\left (\dx \,{\it \us}\right )\dt^{2}u}{1-u{ \it
\us}}}+4\,{\frac {\left (\dx \,u\right )\dt^{2}{\it \us}}{1-u{\it
\us}}}+4\,{\frac {\left (\dt u\right )\dx \dt{\it \us}}{1-u{\it
\us}}} -4\,{\frac {\left (\dt {\it \us}\right )\dx \,\dt
u}{1-u{\it \us}}} \nonumber
\\+4\,{\frac { \left (-u\dx \,{\it \us}+{\it
\us}\,\dx \,u\right )\left ( \left (\dx \,{\it \us}\right )\dx
\,u+\left (\dt u \right )\dt{\it \us}\right )}{\left (1-u{\it
\us}\right )^{2}} } \nonumber \\ +4\,\left (-u\dx \,{\it \us}+{\it
\us}\,\dx \,u\right )\beta \ \ .
\end{eqnarray}
We are looking for boundary conditions that are of the form
\be
 \dx u = F(u,\us) \ \ \ , \ \ \ \dx \us = G(u,\us) \ \ ,
  \ee
where $F$ and $G$ are functions of the fields not involving
derivatives. By direct substitution of the above into
(\ref{ap:l21}) we get
 \begin{eqnarray} \label{ap:l22}
 4\,{\frac {\left (2\, {\it \frac{\d  G}{\d u }}(1 -\,u{\it \us})+G{\it \us}
\right )\left (\dt u\right )^{2}}{\left (1-u{\it \us}\right )^
{2}}} -4\,{\frac {\left (\dt{\it \us} \right )^{2}\left ( 2\,{\it
\frac{\d  F}{\d \us }}(1-\,u{\it \us})+F u\right )} {\left
(1-u{\it \us}\right )^{2}}} \nonumber
\\ 8\,{\frac {\left (\dt
{\it \us}\right )\left (-{\it \frac{\d F}{\d u }}+ {\it \frac{\d
G}{\d \us }}\right )\dt u}{1- u{\it \us}}} +4\,{\frac {\left (-u
G+{\it \us}\,F\right )G F}{\left ( 1-u{\it \us}\right )^{2}}}
\nonumber \\
 +4\,\left (-u G+{\it
\us} \,F\right )\beta \nonumber \ \ .
\end{eqnarray}
The expression above does represent a total derivative when all
terms are forced to vanish by selecting suitable functions $F$ and
$G$. The two separate differential equations that appear involving
the undefined functions
 \begin{eqnarray}
 2 ( 1 - u \us) \frac{ \d F}{ \d \us} + u F &=& 0 \ \ ,\nonumber \\
 2 ( 1 - u \us) \frac{ \d G}{ \d u} + \us G &=& 0 \ \ , \nonumber
 \end{eqnarray}
can easily be solved to yield
 \be \label{ap:l23}
 F(u,\us) = S_1(u) \sqrt { 1 - u \us} \ \ , \ \  G(u,\us) =
 S_2(\us) \sqrt { 1 - u \us} \ \ .
 \ee
In addition, the last two terms in (\ref{ap:l22}) imply that
 \be
F = \frac{u}{\us} \ G \ \ ,
 \ee
Using the above relation and solutions of (\ref{ap:l23}) into the
remaining terms of (\ref{ap:l22}), we can determine the remaining
undefined  functions $S_1$ and $S_2$. The final form of the
boundary conditions are
 \begin{eqnarray} \label{ap:bc}
 \dx u =- C u \sqrt{1 - u \us } \ , \nonumber \\
 \\
 \dx \us =- C \us \sqrt{1 - u \us } \ . \nonumber
 \end{eqnarray}
where C is a real constant.

\bibliographystyle{unsrt}

\end{document}